%% file: ADM.tex
\documentclass{amsart}
\usepackage[margin=1in]{geometry}
\usepackage{amsmath, amsfonts, amssymb}
\usepackage{color}
\usepackage[shortlabels]{enumitem}
\usepackage{caption}
\usepackage{overpic}
\usepackage{subfigure}
\usepackage{mathrsfs}
\usepackage{threeparttable}

\DeclareMathOperator\tr{tr}

\DeclareMathOperator\diag{diag}
\DeclareMathOperator*\argmin{argmin}

\title{Alternating Descent Method for Gauge Cooling of Complex Langevin Simulations}

\author{Xiaoyu Dong}
\address[Xiaoyu Dong]{University of Chinese Academy of Sciences, Beijing 100049, China}
\email{dongxiaoyu@lsec.cc.ac.cn}

\author{Zhenning Cai}
\address[Zhenning Cai]{Department of Mathematics, National University of Singapore,
  Level 4, Block S17, 10 Lower Kent Ridge Road, Singapore 119076}
\email{matcz@nus.edu.sg}
\date{\today}

\author{Yana Di}
\address[Yana Di]{Institute of Mathematical Research, Beijing Normal University \& UIC, Zhuhai 519087, China}
\email{yndi@uic.edu.cn}

\thanks{Zhenning Cai's work was supported by the Academic Research Fund of the Ministry of Education of Singapore under grant No. R-146-000-291-114.
Yana Di's work was supported by National Natural Science Foundation of China No.11771437. The computations were partly done on the high performance computers of State Key Laboratory of Scientific and Engineering Computing, Chinese Academy of Sciences.}

\begin{document}

\begin{abstract}
We study the gauge cooling technique for the complex Langevin method applied to the computation in lattice quantum chromodynamics. We propose a new solver of the minimization problem that optimizes the gauge, which does not include any parameter in each iteration, and shows better performance than the classical gradient descent method especially when the lattice size is large. Two numerical tests are carried out to show the effectiveness of the new algorithm.

\end{abstract}

\maketitle

\input{Introduction.tex}
\input{ComplexLangevin.tex}
\input{GaugeCooling.tex}
\input{Example.tex}

\input{Conclusion.tex}

\bibliographystyle{plain}
\bibliography{NewAlgorithm}
\end{document}

%% file: Introduction.tex
\section{Introduction}
Lattice QCD is the standard nonperturbative tool for quantum chromodynamics (QCD). The link variables $U_{x,\mu} \in SU(3)$ stand for the gluons between lattice points $x$ and $x+\hat{\mu}$ where $x=(t,x_1,x_2,x_3) \in \mathbb{N}^4$. Using $\{U\}$ to represent the collection of all link variables $U_{x,\mu}$, one can compute the expectation value of any observable $O$ by
\begin{equation*}
\langle O \rangle = \frac{1}{Z} \int [\mathrm{d}U] \, O(\{U\}) e^{-S(\{U\})},
\end{equation*}
where the partition function $Z$ is given by
\begin{equation*}
Z = \int [\mathrm{d}U] \, e^{-S(\{U\})},
\end{equation*}
and the total action $S(\{U\})$ can be written as $S_B(\{U\}) + S_F(\{U\})$, with $S_B(\{U\})$ being the bosonic action and $S_F(\{U\})$ being the negative logarithm of the fermion determinant. For nonzero chemical potential, the fermion determinant $\det M$ may be complex, leading to the numerical sign problem when applying Monte Carlo method to compute $\langle O\rangle$.

Several methods have been proposed to mitigate the numerical sign problem, including analytical methods such as analytical continuation and series expansion \cite{Forcrand2002, Allton2003}, the Lefschetz thimble method \cite{Cristoforetti2012, Cristoforetti2013}, and the complex Langevin method (CLM) \cite{Klauder1983, Klauder1983J, Klauder1984, Parisi1983}. The CLM is a straightforward generalization of the real Langevin method, which extends the dynamics on the $SU(3)$ field to the dynamics on the $SL(3,\mathbb{C})$ field. Despite its formal legitimacy \cite{Aarts2010}, the results may diverge or even converge to wrong results \cite{Nagata2016J, Scherzer2019}, and this often occurs when large excursions occur in the complex Langevin dynamics. Recently in \cite{Scherzer2019}, by a detailed analysis on the boundary terms, the underlying mechanism of the wrong convergence result has surfaced, and the result therein is further developed in \cite{Scherzer2020} to find corrections of the numerical values.

Nevertheless, the most straightforward way to stabilize the dynamics is to pull the fields closer to $SU(3)$, which can be done by gauge cooling \cite{Seiler2013} or dynamical stabilization \cite{Attanasio2019}. The gauge cooling method utilizes the redundant degrees of freedom in the gauge field, and it does not introduce any biases to the expectation values. Such a method has been formally justified in \cite{Nagata2016J, Nagata2016}, and it has been successfully applied to a number of problems \cite{Sexty2014, Nagata2016G}. Basically, the gauge cooling method requires to choose an optimal gauge for the complexified field, which minimizes the distance between the current field and $SU(3)$. In one dimension, it has been figured out in \cite{Cai2020} that the problem can be solved analytically since the field is essentially equivalent to the one-link model. While the analytical solution is unavailable for the multi-dimensional case, the minimization problem is usually solved by the gradient descent method, for which the step length has to be chosen carefully to achieve satisfactory convergence rate. Some techniques to choose the step length have been discussed in \cite{Aarts2013, Bloch2018}.

In this paper, we would like to focus on the gauge cooling problem and propose a more efficient method to solve the optimization problem, which does not require delicate selection of the step length. The rest of this paper is organized as follows. In Section \ref{sec:CLM}, we give a brief introduction to the complex Langevin method and the gauge cooling technique. Our alternating descent method is presented in Section \ref{sec:gauge_cooling}. Section \ref{sec:exm} includes two numerical experiments to test the new method. Finally, some concluding remarks are given in Section \ref{sec:conclusion}. 

%% file: ComplexLangevin.tex
\section{Complex Langevin method and gauge cooling} \label{sec:CLM}
In this section, we provide a brief review of the complex Langevim method and the method of gauge cooling.  To be more general, we consider the $SU(n)$ gauge theory with $\{U\} \in [SU(n)]^{4N_0N_1N_2N_3}$, where $N_0$ and $N_1, N_2,N_3$ are the numbers of lattice points in time and spatial directions. Define
\begin{equation*}
G = \{x=(t,x_1,x_2,x_3):\ t=1,\cdots,N_0;\ x_1=1,\cdots,N_1;\ x_2=1,\cdots,N_2;\ x_3=1,\cdots,N_3\},
\end{equation*}
then $\{U\}$ can be represented by $\{U_{x,\mu}: x \in G;\ \mu = 0,1,2,3\}$. We assume periodic boundary conditions for $\{U\}$.

Let $w_{a,x,\mu}$ for $a = 1,\cdots,n^2-1$, $x\in G$ and $\mu = 0,1,2,3$ be independent Brownian motions. The complex Langevin method is described by a complex stochastic process:
\begin{equation} \label{eq:CLE}
\mathrm{d} U_{x,\mu} = -\sum_{a=1}^{n^2-1} \mathrm{i} \lambda_a \left[U_{x,\mu} D_{a,x,\mu}S(\{U\}) \mathrm{d}t + U_{x,\mu} \circ \mathrm{d}w_{a,x,\mu} \right], \qquad \forall x \in G, \quad \forall \mu = 0,1,2,3
\end{equation} 
where $\lambda_a$, $a = 1,\cdots,n^2-1$ are the infinitesimal generators of the $SU(n)$ group satisfying
\begin{equation*}
\tr(\lambda_a \lambda_b) = 2 \delta_{ab}, \qquad \forall a,b=1,\cdots,n^2-1,
\end{equation*}
and $D_{a,x,\mu}$ denotes the left Lie derivative operator defined as
\begin{equation*}
D_{a,x,\mu} f(\{U\}) = \lim_{\epsilon \rightarrow 0} \frac{f(\{U^{\epsilon}\}) - f(\{U\})}{\epsilon}
\end{equation*}
with $\{U^{\epsilon}\}$ being the set of the matrices $U_{y,\nu}^{\epsilon} = \exp(\mathrm{i}\epsilon \delta_{xy} \delta_{\mu\nu} \lambda_a) U_{y,\nu}$ for $y\in G$, $\nu = 0,1,2,3$, $a=1,2,\cdots,n^2-1$. Here the field $\{U^\epsilon\}$ is actually the perturbation of $\{U\}$ which only replaces the single link $U_{x,\mu}$ by $\exp(\mathrm{i \epsilon \lambda_a}) U_{x,\mu}$. In practice, we apply the Euler-Maruyama discretization of the equation (\ref{eq:CLE}) for time evolution:
\begin{equation} \label{eq:Euler}
U_{x,\mu} = \exp \left( -\sum_{a=1}^{n^2-1} \mathrm{i} \lambda_a \left( D_{a,x,\mu}S \Delta t + \eta_{a,x,\mu} \sqrt{\Delta t} \right)  \right) U_{x,\mu}, \quad \forall x \in G, \quad \mu = 0,1,2,3,
\end{equation}
where $\Delta t$ is time step and any of $\eta_{a,x,\mu}$ is normally distributed with mean $0$ and variance $2$.

The complex Langevin method may diverge due to insufficient decay of the probability density function at infinity, which is often owing to too much excursion away from a unitary field. The gauge cooling method uses the redundant degrees of freedom in the gauge theory to restrict such excursion. Specifically, the observable $O$ and the complex action $S$ are invariant under the gauge transformation defined as
\begin{equation} \label{eq:gc}
\tilde{U}_{x,\mu} = V^{-1}_x U_{x,\mu} V_{x+\hat{\mu}}, \qquad \forall x \in G, \quad \forall \mu = 0,1,2,3
\end{equation}
where $V_x \in SL(n,\mathbb{C})$ for any $x \in G$, which also satisfy the periodic boundary conditions. Define $\{\tilde{U}\}$ as the collection of all the $\tilde{U}_{x,\mu}$. The gauge cooling method applies such a transform after every time step. By choosing $V_x$ appropriately, the drift away from the unitary field can be mitigated. The distance between $\{U\}$ and $[SU(n)]^{4N_0 N_1 N_2 N_3}$ can be measured by the norm
\begin{equation*}
\|\{U\}\| =  \left(\frac{1}{4 N_0 N_1 N_2 N_3} \sum_{\mu=0}^3 \sum_{x \in G} \|U_{x,\mu}\|_F^2 \right)^{1/2},\qquad \forall \{U\} \in [\mathbb{C}^{n \times n}]^{4 N_0 N_1 N_2 N_3}
\end{equation*}
based on the Frobenius norm of matrices defined by $\|U_{x,\mu}\|_F = [\tr(U_{x,\mu} U_{x,\mu}^\dagger)]^{1/2}$. It can be proved that $\|\{U\}\| \geqslant n$ for any $\{U\} \in[S L(n, \mathbb{C})]^{4 N_0 N_1 N_2 N_3}$, and the equality holds if and only if $\{U\} \in[S U(n)]^{4 N_0 N_1 N_2 N_3}$ \cite{Cai2020}.

Once the link variables $\{U\}$ are updated by one time step, the gauge cooling step requires us to solve the minimization problem:
\begin{equation} \label{eq:min}
\argmin_{\{\tilde{U}\} \in \mathcal{G}(\{U\})} F(\{\tilde{U}\})
\end{equation}
where $F(\{\tilde{U}\}) = \|\{\tilde{U}\}\|^2$, and $\mathcal{G}(\{U\})$ is the set of all fields which are equivalent to $\{U\}$ under the gauge transformation:
\begin{equation*}
\mathcal{G}(\{U\}) = \left\{ \{\tilde{U}\} :
  \tilde{U}_{x,\mu} = V^{-1}_x U_{x,\mu} V_{x+\hat{\mu}}, \, V_x \in SL(n,\mathbb{C}) \right\}.
\end{equation*}
In \cite{Seiler2013}, this problem is solved iteratively by the gradient descent method:
\begin{equation} \label{eq:gd}
\tilde{U}_{x,\mu}^{(0)} = U_{x,\mu}, \qquad
\tilde{U}_{x,\mu}^{(k+1)} = \exp \left(\sum_{a=1}^{n^2-1} s V_{a,x}^{(k)} \lambda_a \right) \tilde{U}_{x,\mu}^{(k)} \exp \left(-\sum_{a=1}^{n^2-1} s V_{a,x+\hat{\mu}}^{(k)} \lambda_a \right),
\end{equation}
where
\begin{equation*}
V_{a,x}^{(k)} = - 2 \sum_{\mu=0}^3 \tr \lambda_a \left[ \tilde{U}_{x,\mu}^{(k)} (\tilde{U}_{x,\mu}^{(k)})^\dagger - (\tilde{U}_{x-\hat{\mu},\mu}^{(k)})^\dagger \tilde{U}_{x-\hat{\mu},\mu}^{(k)} \right],
\end{equation*}
and $s$ stands for the step length. Since the optimization problem has to be solved at every time step, an accurate line search method may be too expensive for gauge cooling. The simplest method is to choose $s$ to be proportional to $\Delta t$ \cite{Seiler2013}, while it is unclear how to select a suitable coefficient. In \cite{Aarts2013}, the authors introduced the method of adaptive gauge cooling, which sets $s$ to be larger when the current solution is farther away from the minimum point, and controls the divergence by reducing $s$ when the cooling drift is large. In \cite{Bloch2018}, the authors find out $s$ numerically by applying Brent's method \cite{Brent1971} and adding an upper bound to avoid instability.

%% file: GaugeCooling.tex
\section{Alternating descent method for gauge cooling} \label{sec:gauge_cooling}
To avoid intricate selection of the time step, we will introduce in this section a new numerical method of gauge cooling, which is free of parameters in each iteration. The general idea of this method is to decompose all the variables $V_x$ into two subsets, and these two subsets of variables will be optimized in turns. Below we refer to this method as the \emph{alternating descent method}.

To begin with, we define two index sets:
\begin{equation*}
\begin{aligned}
& G_e = \{x=(t,x_1,x_2,x_3) \in G: (t+x_1+x_2+x_3) \ \mathrm{is} \ \mathrm{even}\}, \\
& G_o = \{x=(t,x_1,x_2,x_3) \in G: (t+x_1+x_2+x_3) \ \mathrm{is} \ \mathrm{odd}\}.
\end{aligned}
\end{equation*}
In order to be consistent with the periodic boundary condition, we require that $N_0, N_1, N_2$ and $N_3$ be even numbers. Each iteration in the alternating descent method then contains two steps:
\begin{enumerate}
\item Let $\tilde{U}_{x,\mu}^{{(k+1/2)}} = (V_x^{(k)})^{-1} \tilde{U}_{x,\mu}^{(k)}$ and $\tilde{U}_{x+\hat{\mu},\mu}^{(k+1/2)} = \tilde{U}_{x+\hat{\mu},\mu}^{(k)} V_x^{(k)}$ for all $x \in G_e$, where $V_x^{(k)}$, $x\in G_e$ are chosen to minimize $F(\{\tilde{U}^{(k+1/2)}\})$.
\item Let $\tilde{U}_{x,\mu}^{(k+1)} = (V_x^{(k+1/2)})^{-1} \tilde{U}_{x,\mu}^{(k+1/2)}$ and $\tilde{U}_{x+\hat{\mu},\mu}^{(k+1)} = \tilde{U}_{x,\mu}^{(k+1/2)} V_x^{(k+1/2)}$ for all $x \in G_o$, where $V_x^{(k+1/2)}$, $x\in G_o$ are chosen to minimize $F(\{\tilde{U}^{(k+1)}\})$.
\end{enumerate}
The operation in each step finds the optimal gauge transformation with transformation variables $V_x^{(k)}$ (or $V_x^{(k+1/2)}$) being fixed to be identity on odd (or even) lattice points. To implement the above algorithm, it remains only to find $V_x^{(k)}$ for $x \in G_e$ and $V_x^{(k+1/2)}$ for $x \in G_o$. Due to the periodic boundary condition, the algorithms for finding both sets of matrices are nearly identical, and below we focus only on the algorithm to find $V_x^{(k)} \in G_e$. For simplicity, we omit the superscript $k$, and thus the optimization problem to be solved is
%
%
\begin{equation} \label{eq:norm_e}
\argmin_{V_x \in SL(n), \, x\in G_e} \sum_{\mu=0}^3 \sum_{x \in G_e} \tr(V_x^{-1} \tilde{U}_{x,\mu} \tilde{U}_{x,\mu}^\dagger V_x^{-\dagger} + \tilde{U}_{x-\hat{\mu},\mu} V_x V_x^\dagger \tilde{U}_{x-\hat{\mu},\mu}^\dagger).
\end{equation}
Note that the objective function is exactly the value of $F(\cdot)$ for link variables after the gauge transformation. In this problem, the optimization of every $V_x$ can be fully decoupled, meaning that we can solve the minimization problem
\begin{equation} \label{eq:opt}
\argmin_{V_x \in SL(n)} \mathscr{F}_x(V_x) = \sum_{\mu=0}^3 \tr(V_x^{-1} \tilde{U}_{x,\mu} \tilde{U}_{x,\mu}^\dagger V_x^{-\dagger} + \tilde{U}_{x-\hat{\mu},\mu} V_x V_x^\dagger \tilde{U}_{x-\hat{\mu},\mu}^\dagger)
\end{equation}
for all $x \in G_e$, as makes it feasible to solve \eqref{eq:norm_e} exactly.
Now we fix $x \in G_e$ and try to solve \eqref{eq:opt}. Since $V_x \in SL(n,\mathbb{C})$, we can write $V_x$ as
\begin{equation*}
V_{x} = \exp \left(-\sum_{a=1}^{n^2-1} V_{a,x} \lambda_a \right),
\end{equation*}
where all $V_{a,x}$ are real numbers as in the gradient descent method. By straightforward calculation, we have
\begin{equation} \label{eq:grad}
\frac{\partial \mathscr{F}_x}{\partial V_{a,x}} = 2 \tr \left[\lambda_{a} \sum_{\mu=0}^3 (V_x^{-1} \tilde{U}_{x, \mu} \tilde{U}_{x, \mu}^\dagger V_x^{-\dagger} - V_x^\dagger \tilde{U}_{x-\hat{\mu}, \mu}^\dagger \tilde{U}_{x-\hat{\mu}, \mu} V_x)\right].
\end{equation}
For simplicity, we let
\begin{equation*}
P_x = \sum_{\mu=0}^3 \tilde{U}_{x, \mu} \tilde{U}_{x, \mu}^\dagger, \qquad
Q_x = \sum_{\mu=0}^3 \tilde{U}_{x-\hat{\mu}, \mu}^\dagger \tilde{U}_{x-\hat{\mu}, \mu},
\end{equation*}
so that the optimal choice of $V_{a,x}$ satisfies
\begin{equation*}
\tr \left[ \lambda_{a} \left( V_x^{-1} P_x V_x^{-\dagger}-V_x^\dagger Q_x V_x \right) \right] = 0, \quad \forall a = 1,\cdots, n^2-1.
\end{equation*}
Since the matrix $V_x^{-1} P_x V_x^{-\dagger} - V_x^\dagger Q_x V_x$ is Hermitian, the above equation is equivalent to
\begin{equation} \label{eq:opt_con2}
V_x^{-1} P_x V_x^{-\dagger}-V_x^\dagger Q_x V_x = \alpha_x I,
\end{equation}
where $\alpha_x \in \mathbb{R}$ is an unknown coefficient.

Now we further simplify the notation by omitting the variable $x$, so that the equation \eqref{eq:opt_con2} can be rewritten as
\begin{equation}
V V^\dagger Q V V^\dagger + \alpha V V^\dagger = P.
\end{equation}
If $\alpha$ is given, this equation is the algebraic Riccati equation \cite{Lancaster1995}. The standard way to solve this equation is to first solve the eigenvalue problem
\begin{equation} \label{eq:ev}
\begin{pmatrix}
\frac{1}{2} \alpha I & Q \\
P & -\frac{1}{2} \alpha I
\end{pmatrix}
\begin{pmatrix} X \\ Y \end{pmatrix} =
\begin{pmatrix} X \\ Y \end{pmatrix} \Lambda,
\end{equation}
where $X,Y,\Lambda \in \mathbb{C}^{n \times n}$ and $\Lambda$ is a diagonal matrix whose all diagonal elements have positive real parts. Then the solution can be written as $V V^\dagger = Y X^{-1}$. Here $\Lambda$ only gives a half of the eigenvalues of the first matrix in \eqref{eq:ev}, and the other half of the eigenvalues must have negative real parts. The solution contains the parameter $\alpha$, which can be further determined by $\det(V V^\dagger) = 1$. In our case, we can simplify the problem by expanding \eqref{eq:ev}:
\begin{equation} \label{eq:eig_Z}
\frac{1}{2} \alpha X + Q Y = X \Lambda, \quad P X-\frac{1}{2} \alpha Y = Y \Lambda,
\end{equation}
which yields
\begin{equation}  \label{eq:eig_decom}
Q P X = X \left(\Lambda - \frac{1}{2} \alpha I\right) \left(\Lambda + \frac{1}{2} \alpha I\right)
 = X \left(\Lambda^{2}-\frac{1}{4} \alpha^2 I \right).
\end{equation}
Since $P$ and $Q$ are Hermitian positive definite matrices, their product matrix $QP$ is diagonalizable and all the eigenvalues are real. Thus $X$ can be obtained by solving a smaller eigenvalue problem \eqref{eq:eig_decom}. Here we assume that $\det X = 1$. Suppose $\xi_1, \cdots, \xi_n$ are the eigenvalues of $QP$. Then
\begin{displaymath}
\Lambda = \diag \left(\sqrt{\xi_1 + 1/4 \alpha^2}, \sqrt{\xi_2 + 1/4 \alpha^2}, \cdots, \sqrt{\xi_n + 1/4 \alpha^2} \right).
\end{displaymath}
By the second equation in \eqref{eq:eig_Z}, we have
\begin{displaymath}
Y = P X \left( \Lambda + \frac{1}{2} \alpha I \right)^{-1}.
\end{displaymath}
Therefore
\begin{equation} \label{eq:VVd}
V V^{\dagger} = Y X^{-1} = P X \left( \Lambda + \frac{1}{2} \alpha I \right)^{-1} X^{-1}.
\end{equation}
Using the condition $\det(V) = 1$ and $\det(X) = 1$, we see that
\begin{displaymath}
\frac{\det (P)} {\det\left( \Lambda + \frac{1}{2} \alpha I \right)} = 1.
\end{displaymath}
Therefore $\alpha$ satisfies
\begin{equation} \label{eq:alpha}
\prod_{j=1}^n \left(\sqrt{\xi_j + 1/4 \alpha^2} + 1/2 \alpha \right) = \det(P).
\end{equation}
Note that this equation determines a unique solution, due to the monotonicity of the left-hand side with respect to $\alpha$.
Thus we obtain the following algorithm to solve $V$:
\begin{itemize}
\item[1)] Eigendecompose the matrix $QP$ to get the eigenvalues $\xi_1, \xi_2,\cdots, \xi_n$ and eigenvectors $X$ with $\det(X) = 1$. 

\item[2)] Solve the equation \eqref{eq:alpha} to find the value of $\alpha$:

\item[3)] Compute $V V^{\dagger}$ by \eqref{eq:VVd}.

\item[4)] Find $V$ by taking the matrix square root of \eqref{eq:VVd}.
\end{itemize}
The above algorithm solves the optimization problem \eqref{eq:opt} exactly, and thus provides the solution of \eqref{eq:norm_e}.

This method turns out to be similar to the coordinate descent method, which updates one variable at a time. The coordinate descent method has wide applications in large-scale optimization problems \cite{Nesterov2012}, especially when the subproblem for each variable can be solved exactly \cite{Hsiesh2008, Wang2019}. In the following section, we will show by numerical tests that such a method is also highly efficient in the gauge cooling problem.

%% file: Example.tex
\section{Numerical Examples} \label{sec:exm}
We will now apply the alternating descent method to two numerical examples to demonstrate its efficiency. In both examples, the Euler-Maruyama method \eqref{eq:Euler} is applied to solve the complex Langevin dynamics. Our method will be compared with the gradient descent method, and when the alternating descent method is used, only one iteration is applied after each time step. The two examples to be shown below are, respectively, a one-dimensional Polyakov loop model and a four-dimensional model for heavy quark QCD.

\subsection{Polyakov loop model} \label{sec:1D}
For the one-dimensional Polyakov loop model, We denote the link variables by $U_k, k = 1,2,\cdots,N$. Following \cite{Seiler2013}, we define the action $S$ by
\begin{equation*}
S(\{U\}) = -\tr(\beta_1 U_1 \cdots U_N + \beta_2 U_N^{-1} \cdots U_1^{-1}),
\end{equation*}
and choose $\beta_1 = \beta + \kappa e^\mu$ and $\beta_2 = \bar{\beta} + \kappa e^{-\mu}$ with $\beta = 2, \kappa=0.1$ and $\mu = 1$. The model is applied to the $SU(3)$ theory, so that $\lambda_a$, $a = 1,\cdots,8$ are Gell-Mann matrices. For the one-dimensional model, the parameter $\alpha_x$ in \eqref{eq:opt_con2} always equals zero since both $V_x^{-1} P_x V_x^{-\dagger}$ and $V_x^\dagger Q_x V_x$ are Hermitian, positive definite, and have the determinant $1$.

In our numerical experiments, we choose the Langevin time step to be $\Delta t = 2\times 10^{-5}$ and the simulation ends at $t=10$. Note that in the one-dimensional case, the exact solution of the optimization problem \eqref{eq:min} can be obtained analytically, and the result is
\begin{equation*}
U_k \leftarrow (\Lambda \Lambda^\dagger)^{\frac{1}{2N}}, \quad k=1,\cdots,N-1;  \qquad  U_N \leftarrow \Lambda (\Lambda \Lambda^\dagger)^{-\frac{N-1}{2N}},
\end{equation*}
where $\Lambda$ is a diagonal matrix whose diagonal entries are all the eigenvalues of $U_N U_1 \cdots U_{N-1}$. For details, we refer the readers to a recent work \cite{Cai2020}. Thus, the following four methods will be tested in this example:
\begin{itemize}
\item Complex Langevin with no gauge cooling.
\item Complex Langevin with gauge cooling implemented by the gradient descent method \cite{Seiler2013}.
\item Complex Langevin with optimal gauge cooling \cite{Cai2020}.
\item Complex Langevin with gauge cooling implemented by the alternating descent method introduced in Section \ref{sec:gauge_cooling}.
\end{itemize}
For the gradient descent method, the update of the link variables (\ref{eq:gd}) can be formulated by
\begin{equation*}
V_{a,k} \leftarrow - 2 \tr \lambda_a \left[ U_k U_k^\dagger - U_{k-1}^\dagger U_{k-1}\right],\quad
U_k \leftarrow \exp \left(\sum_{a=1}^8 s V_{a,k} \lambda_a \right) U_k \exp \left(-\sum_{a=1}^8 s V_{a,k+1} \lambda_a \right),
\end{equation*}
where $s$ stands for the step length which is set as $s = \Delta t$. Such iteration is applied three times after every Langevin step. For the alternating descent method, only one iteration per time step is applied as mentioned in the beginning of this section.

\begin{figure}
  \centering
  \subfigure{
    \includegraphics[width=2.5in]{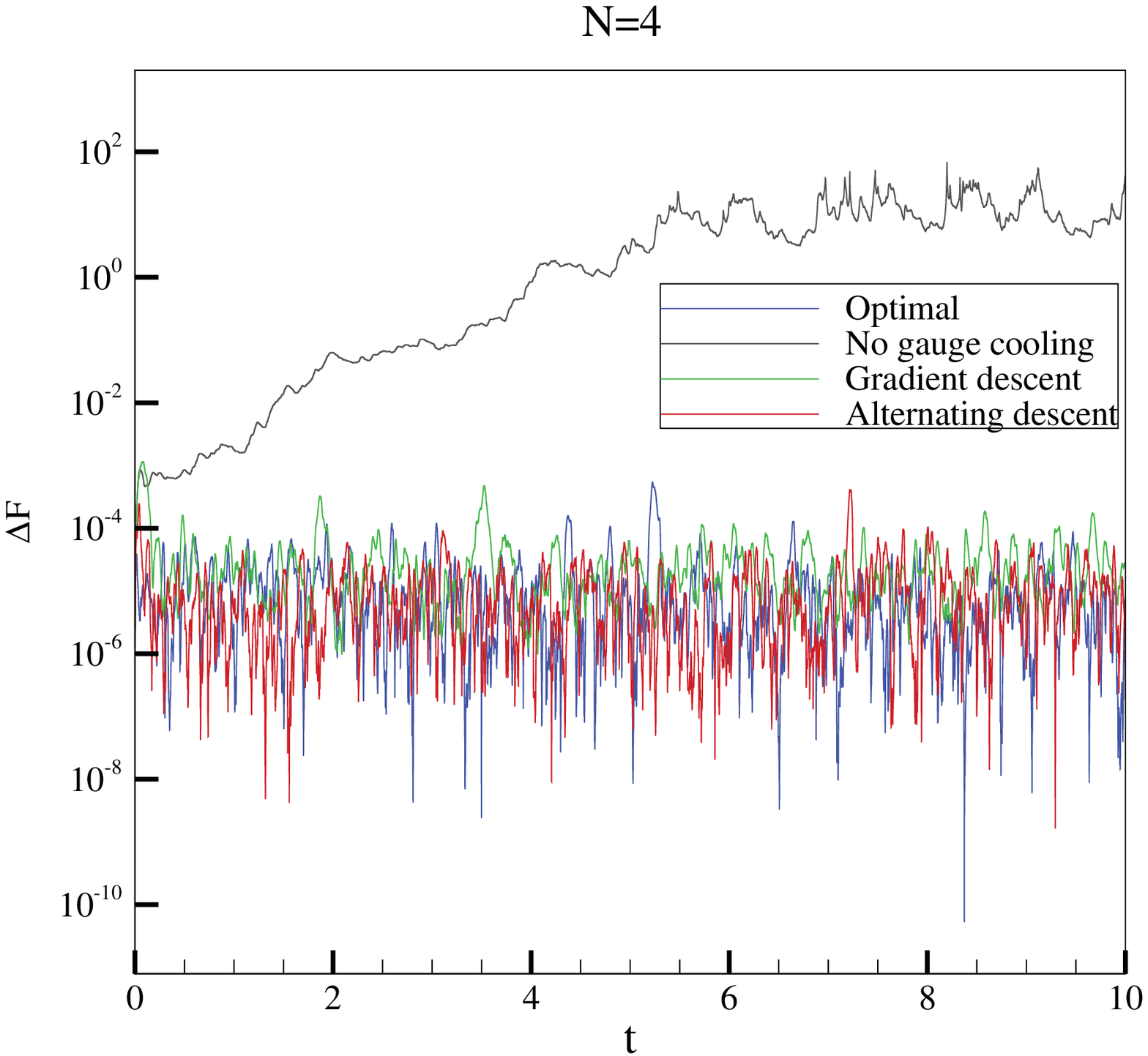}}
  \subfigure{
    \includegraphics[width=2.5in]{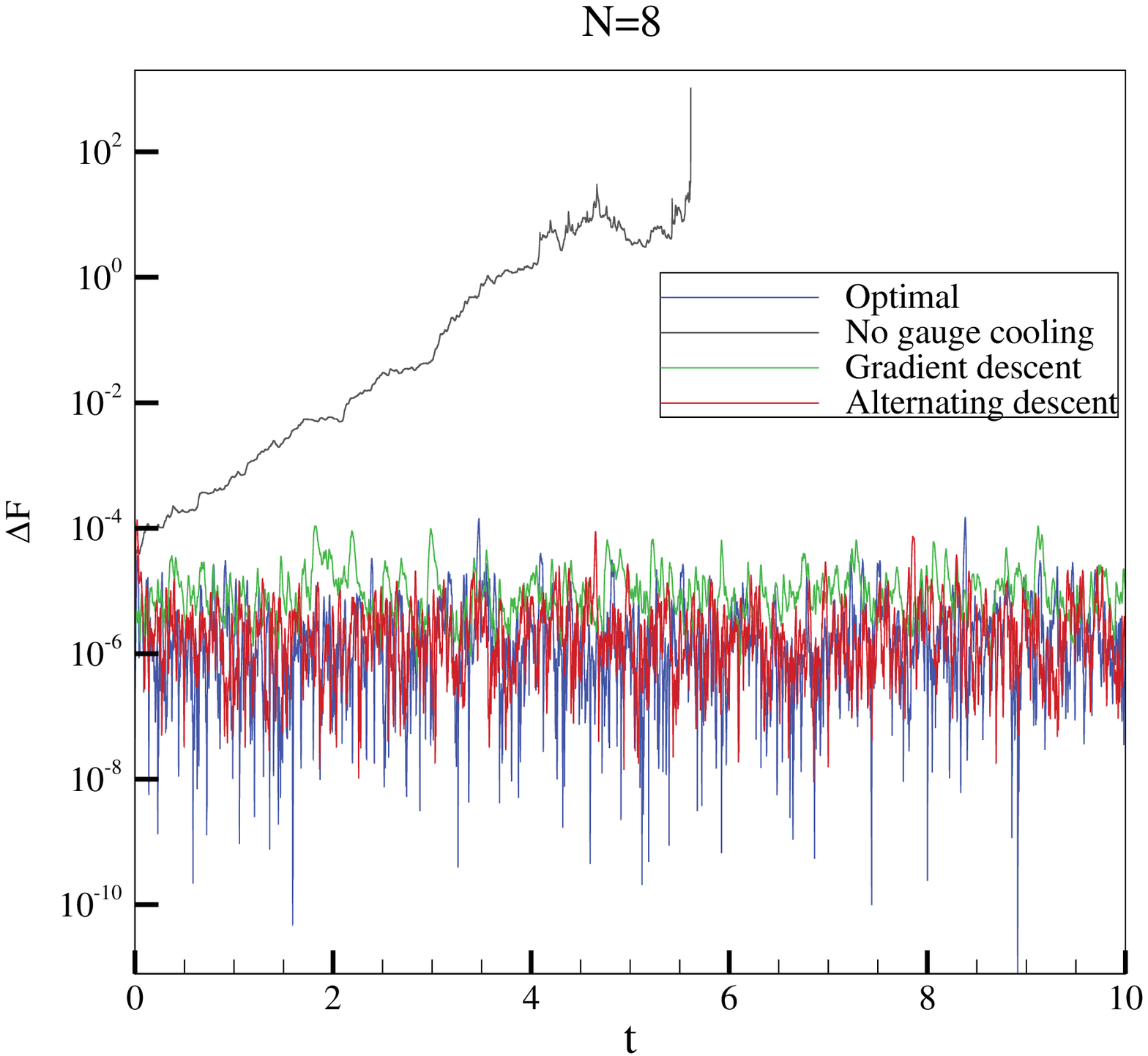}} \\
  \subfigure{
    \includegraphics[width=2.5in]{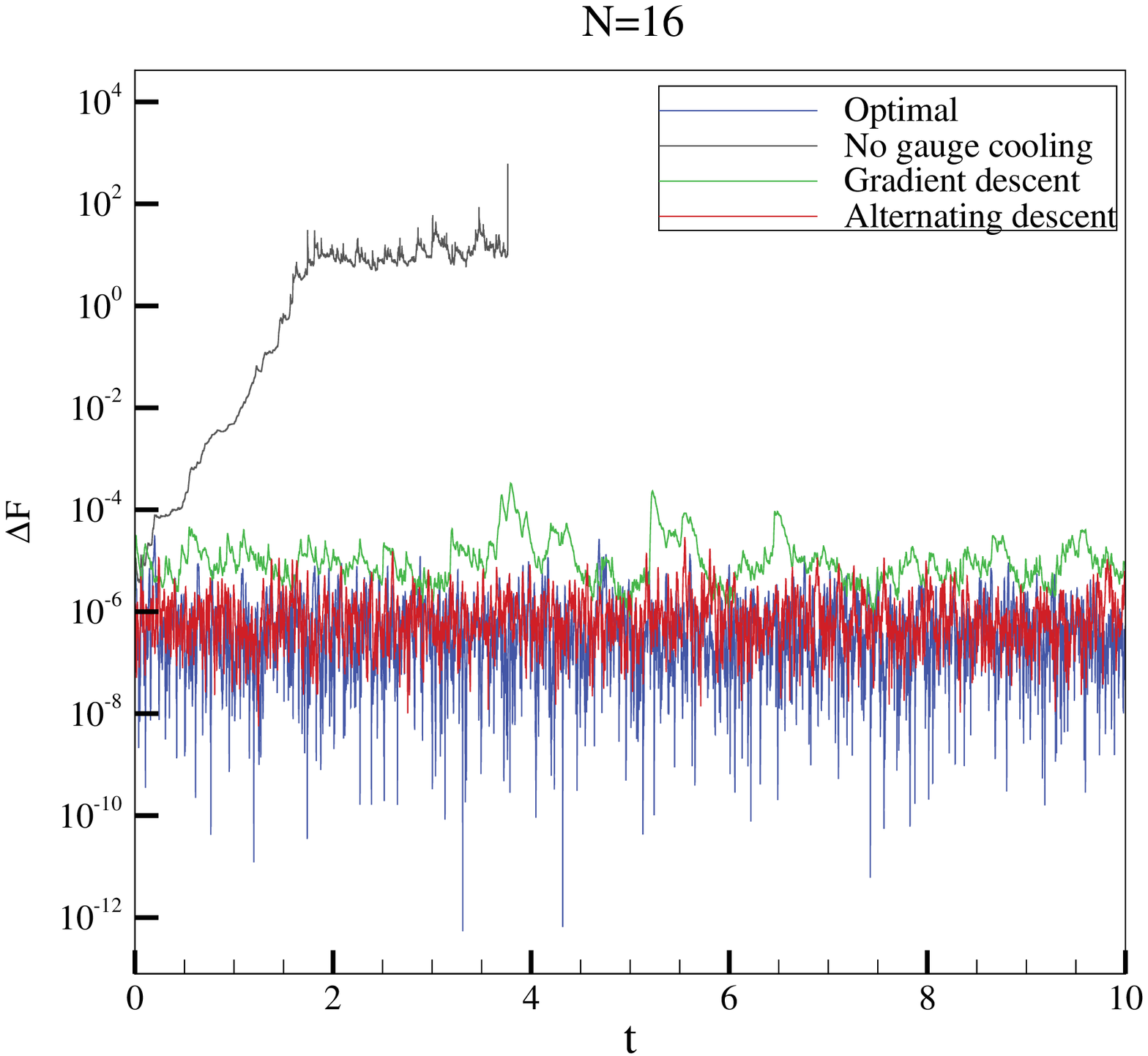}}
  \subfigure{
    \includegraphics[width=2.5in]{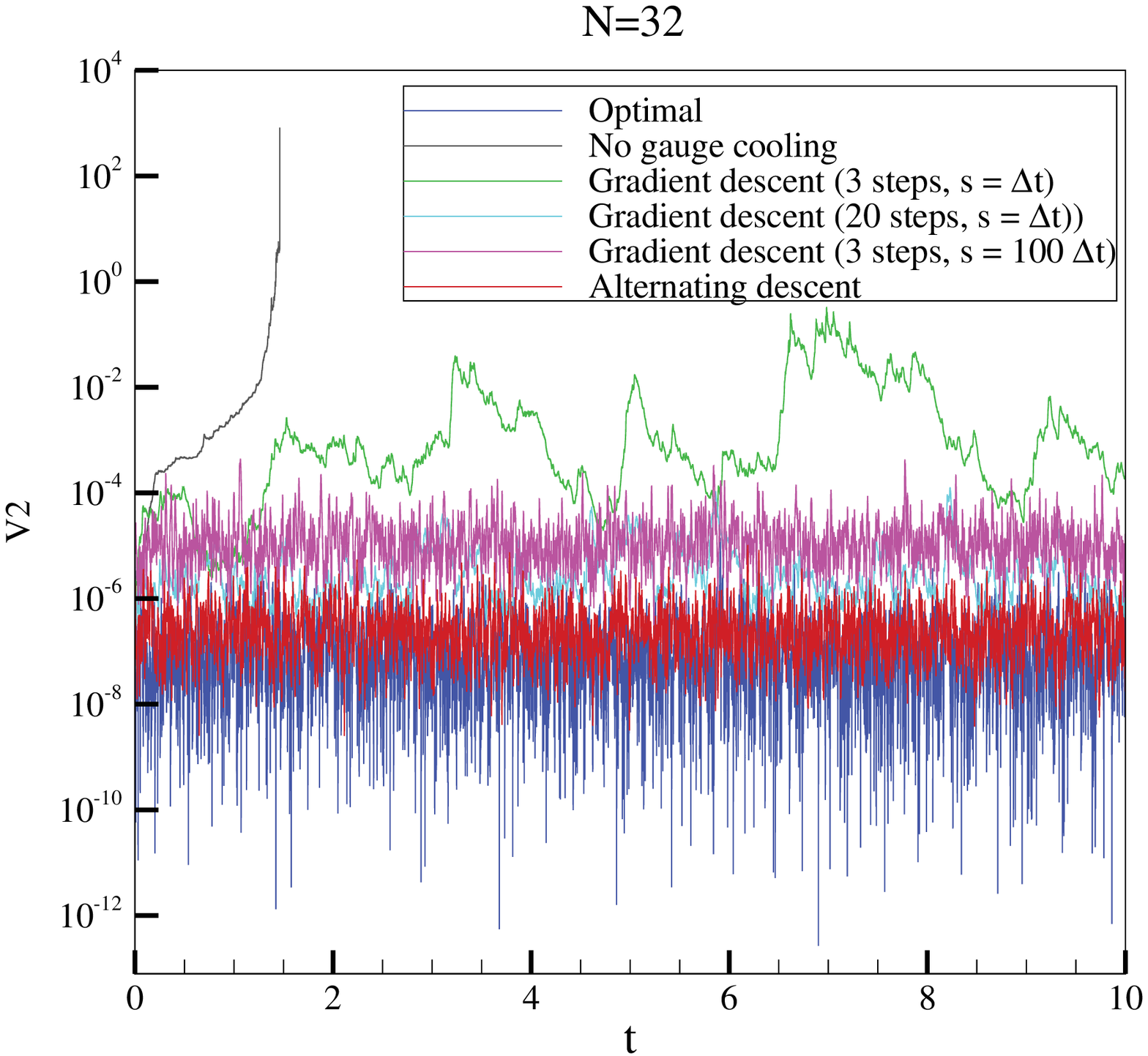}}
  \caption{The evolution of $\Delta F$ with different gauge cooling techniques.}
  \label{fig:norm_1} 
\end{figure} 
Fig.\ref{fig:norm_1} shows the evolution of norm $\|\{U\}\|$ over the Langevin time. The vertical axis $\Delta F$ is defined by $\Delta F = F(\{U\})- 3$, which vanishes only if $\{U\} \in [SU(3)]^N$. Because of the randomness, each curve in the figures only show the result of one particular run in our simulations. From the figures, we can observe without surprise that if no gauge cooling is applied, the computations are forced to terminate for $N = 8,16,32$ due to overflow, and even for $N=4$, the norm of the links increases too quickly. For $N=4$ and $8$, the performance of gradient descent method is excellent due to its good agreement with the optimal gauge cooling. But the disparity between these two methods increases gradually as $N$ gets larger, as is obvious for $N=16$ and $32$. On the contrary, in all cases, the alternating descent method performs almost as well as the optimal gauge cooling. In general, we see that the alternating descent method is always competitive, and when the number of lattice points are large, the alternating descent method can significantly outperform the gradient descent method. To demonstrate the efficiency of the method, we have also listed in Table \ref{tab:cost1} the computational time in our experiments. In Table \ref{tab:cost1}, the total computational time and the computational time spent only on gauge cooling are both provided, where for the gradient descent method, three iterations are applied at every time step, while only one iteration per time step is applied for the alternating descent method. Averagely, the computational time for every iteration is similar for both methods, and the alternating descent method is faster due to its capability of using less iterations. By comparing the two columns for each method, it can be seen that gauge cooling takes a significant portion of total computational time, so that a better algorithm for optimization can effectively reduce the computational cost.

\begin{table}[!ht]
\centering 
  \caption{Computational time (in seconds) of every $10^4$ Langevin steps with gauge cooling.  Opt.: Optimal gauge cooling, GD: Gradient descent method, AD: Alternating descent method, g.c.: gauge cooling}
\label{tab:cost1}
\begin{tabular}{cccccccccc}
\hline \hline 
& & \multicolumn{2}{c}{Opt.} & & \multicolumn{2}{c}{GD} & & \multicolumn{2}{c}{AD}  \\
\hline
$N$ & & total & g.c. & & total & g.c. & & total & g.c.\\
\hline
$4$ & & $0.598$ & $0.061$ & & $1.441$ & $0.890$ & & $0.868$ & $0.314$ 
\\
$8$ & & $1.157$ & $0.071$ & & $2.925$ & $1.787$ & & $1.745$ & $0.623$ 
\\
$16$ & & $2.342$ & $0.087$ & & $5.889$ & $3.551$ & & $3.585$ & $1.242$ 
\\
$32$ & & $5.044$ & $0.119$ & & $12.280$ & $7.151$ & & $7.619$ & $2.553$ \\
\hline \hline
\end{tabular} 
\end{table}

It is also illustrated in Fig. \ref{fig:norm_1} that for the gradient descent method applied to $N=32$, if we increase the number of iterations to 20, the dynamics can be better stablized, although the value of $\Delta F$ is still about one-magnitude larger than the alternating descent method. However, this also results in a significant increment of the computational cost. By tuning the parameters, we find that choosing the time step $s = 100 \Delta t$ and keeping the number of iterations to be $3$ can provide a better balance between the computational cost and the efficacy of gauge cooling. Even so, the alternating descent method is still superior in this case.

To verify the validity of these numerical methods, we calculated the expectation values of observables $O_k(\{U\}) = \tr [(U_1 U_2 \cdots U_N)^k]$ for $k = \pm 1, \pm 2 \pm 3$. Samples are taken every $50$ steps until $t=10$ after $t=1$. The results are listed in Table \ref{tab:observable}, and because the value of $\langle O_k \rangle$ is real, its imaginary part is omitted. The exact values of the expectation can be obtained by Weyl's integration theorem. For $N=4$, although the expectation values can be calculated without gauge cooling, the results are obviously erroneous. For $N=4,8$ and $16$, all the gauge cooling methods give reasonable approximations of the exact values. When $N=32$, the data for gradient descent method provided in Table \ref{tab:observable} are computed using $s = \Delta t$ with 3 iterations per step, which show large errors especially when $|k|$ is large.

\begin{table}[!ht]
\centering \small
\caption{Numerical results for the observables $\langle O_k \rangle$.
  Opt.: Optimal gauge cooling, GD: Gradient descent method, AD: Alternating descent method, No g.c.: No gauge cooling}
\label{tab:observable}
\begin{tabular}{cc@{\hspace{30pt}}cccccc@{\hspace{30pt}}cccccc}
\hline \hline 
& & & \multicolumn{4}{c}{$N=4$} & & \multicolumn{4}{c}{$N=8$} \\
\hline
$k$ & Exact & & Opt. & GD & AD & No g.c. & & Opt. & GD & AD & No g.c.\\
\hline
$1$ & $2.0957$ & & $2.0913$ & $2.1028$ & $2.1051$ & $2.4030$ & & $2.0961$ & $2.0894$ & $2.1006$ & -\\
$-1$ & $2.1026$ & & $2.0984$ & $2.1096$ & $2.1117$ & $2.4679$ & & $2.1031$ & $2.0964$ & $2.1070$ & - \\
$2$ & $0.3761$ & & $0.3573$ & $0.3885$ & $0.3953$ & $-0.4699$ & & $0.3824$ & $0.3604$ & $0.3663$ & - \\
$-2$ & $0.4092$ & & $0.3904$ & $0.4217$ & $0.4273$ & $1.6144$ & & $0.4151$ & $0.3940$ & $0.3979$ & - \\
$3$ & $-0.5269$ & & $-0.5539$ & $-0.5208$ & $-0.5223$ & $-21.6151$ & & $-0.5251$ & $-0.5349$ & $-0.5673$ & - \\
$-3$ & $-0.4800$ & & $-0.5079$ & $-0.4730$ & $-0.4753$ & $-38.8656$ & & $-0.4785$ & $-0.4875$ & $-0.5215$ & - \\
\hline \hline
& & & \multicolumn{4}{c}{$N=16$} & & \multicolumn{4}{c}{$N=32$} \\
\hline
$k$ & Exact & & Opt. & GD & AD & No g.c. & & Opt. & GD & AD & No g.c.\\
\hline
$1$ & $2.0957$ & & $2.0854$ & $2.0885$ & $2.0940$ & - & & $2.0926$ & $2.1082$ & $2.0808$ & - \\
$-1$ & $2.1026$ & & $2.0924$ & $2.0953$ & $2.1009$ & - & & $2.0996$ & $2.1150$ & $2.0879$ & - \\
$2$ & $0.3761$ & & $0.3582$ & $0.3533$ & $0.3688$ & - & & $0.3760$ & $0.3476$ & $0.3534$ & - \\
$-2$ & $0.4092$ & & $0.3923$ & $0.3862$ & $0.4021$ & - & & $0.4094$ & $0.3882$ & $0.3874$ & - \\
$3$ & $-0.5269$ & & $-0.5216$ & $-0.5528$ & $-0.5420$ & - & & $-0.5110$ & $-2.5277$ & $-0.5170$ & -  \\
$-3$ & $-0.4800$ & & $-0.4741$ & $-0.5062$ & $-0.4939$ & - & & $-0.4642$ & $-3.8793$ & $-0.4697$ & -  \\
\hline \hline
\end{tabular} 
\end{table}

\begin{figure}
  \centering
  \subfigure{
    \includegraphics[width=2.5in]{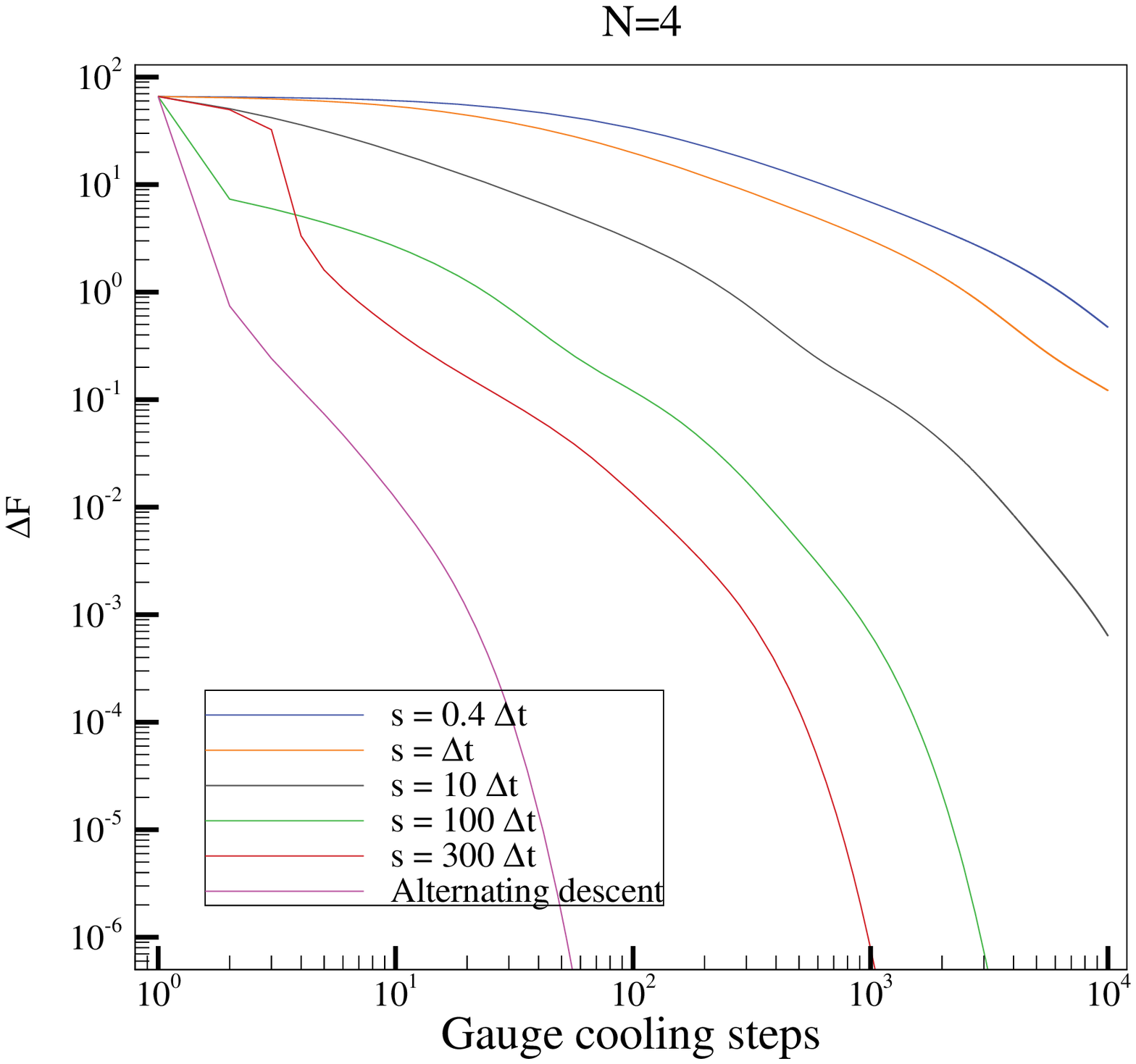}}
  \subfigure{
    \includegraphics[width=2.5in]{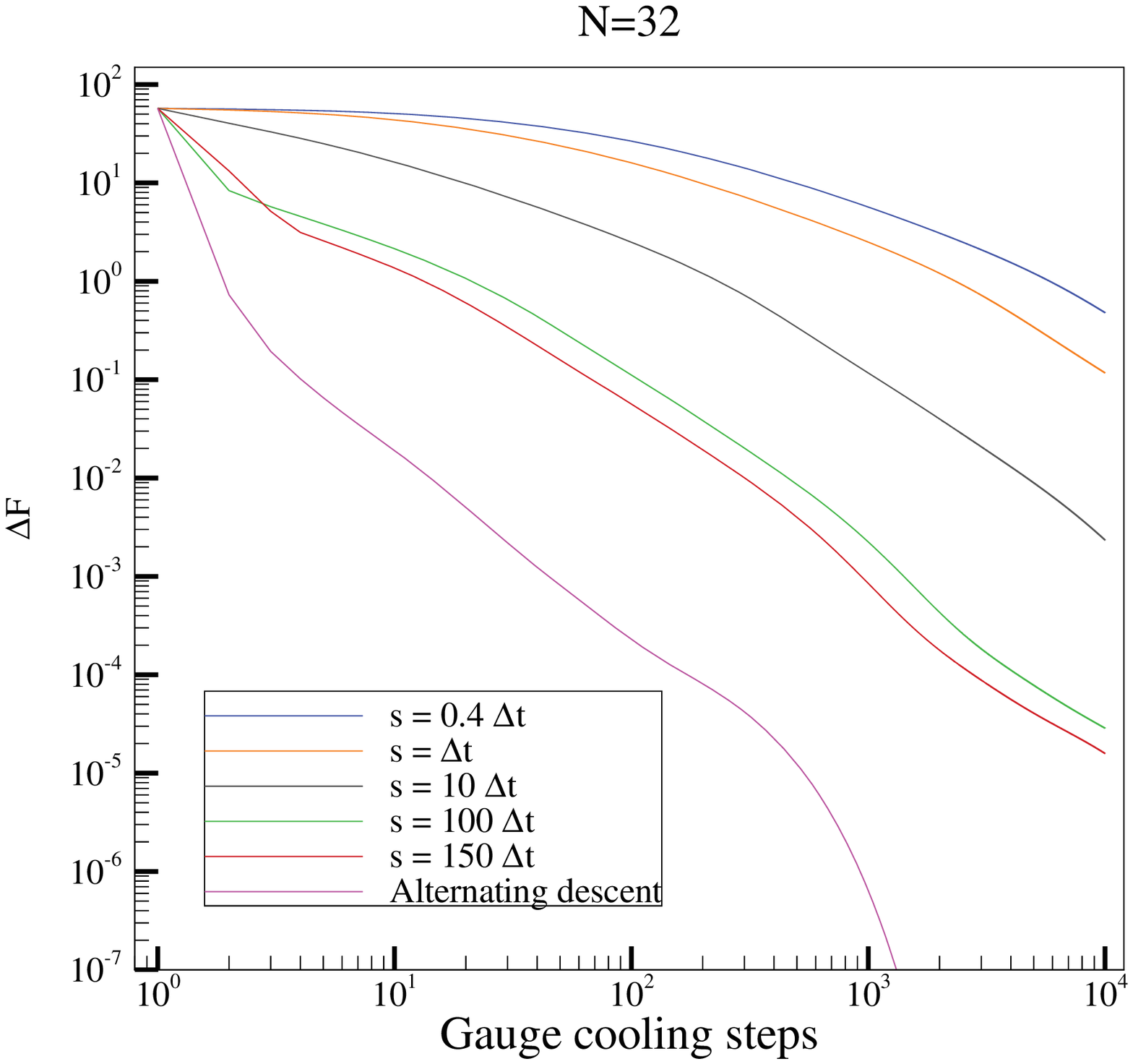}} \\
  \subfigure{
    \includegraphics[width=2.5in]{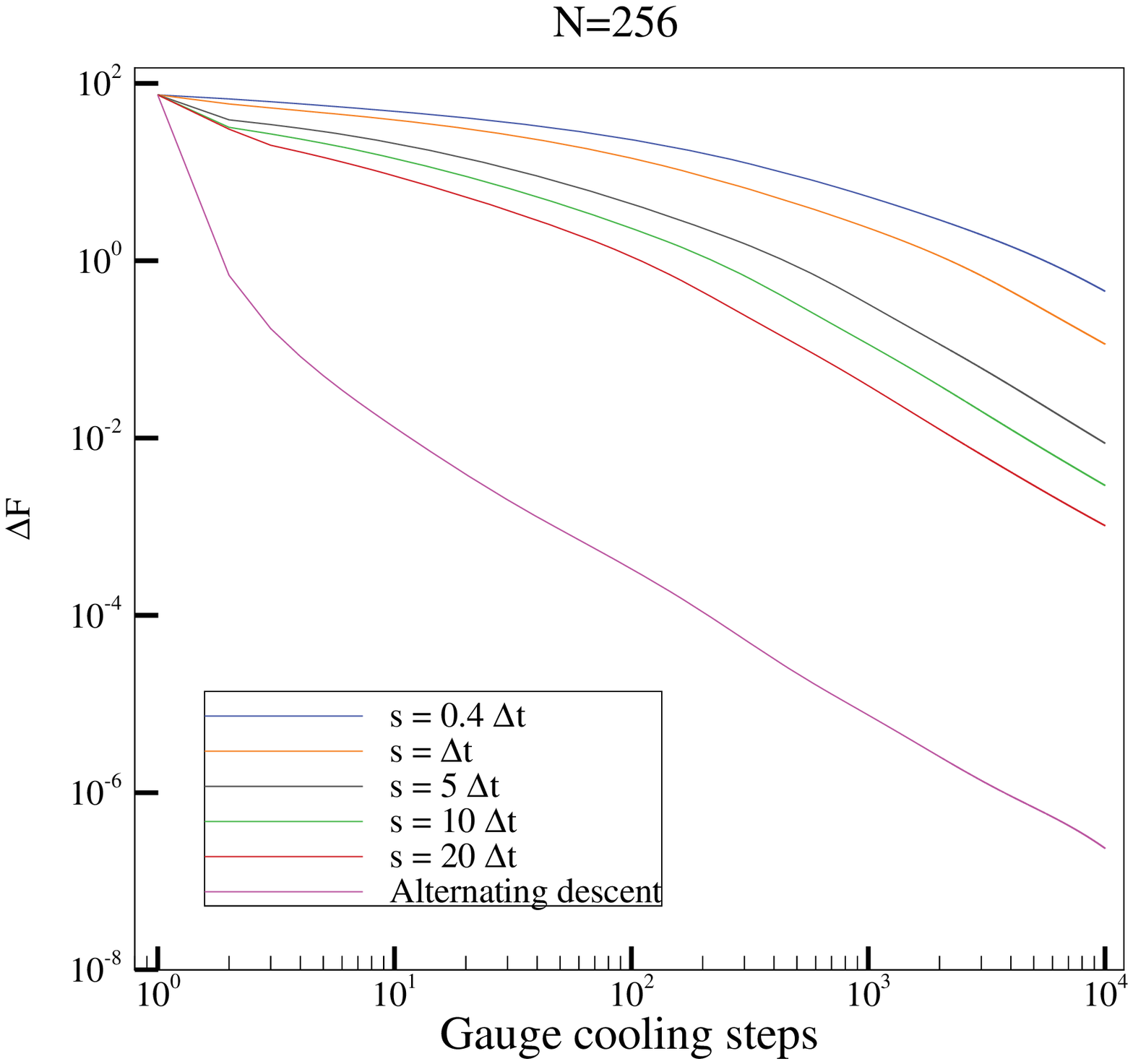}}
  \subfigure{
    \includegraphics[width=2.5in]{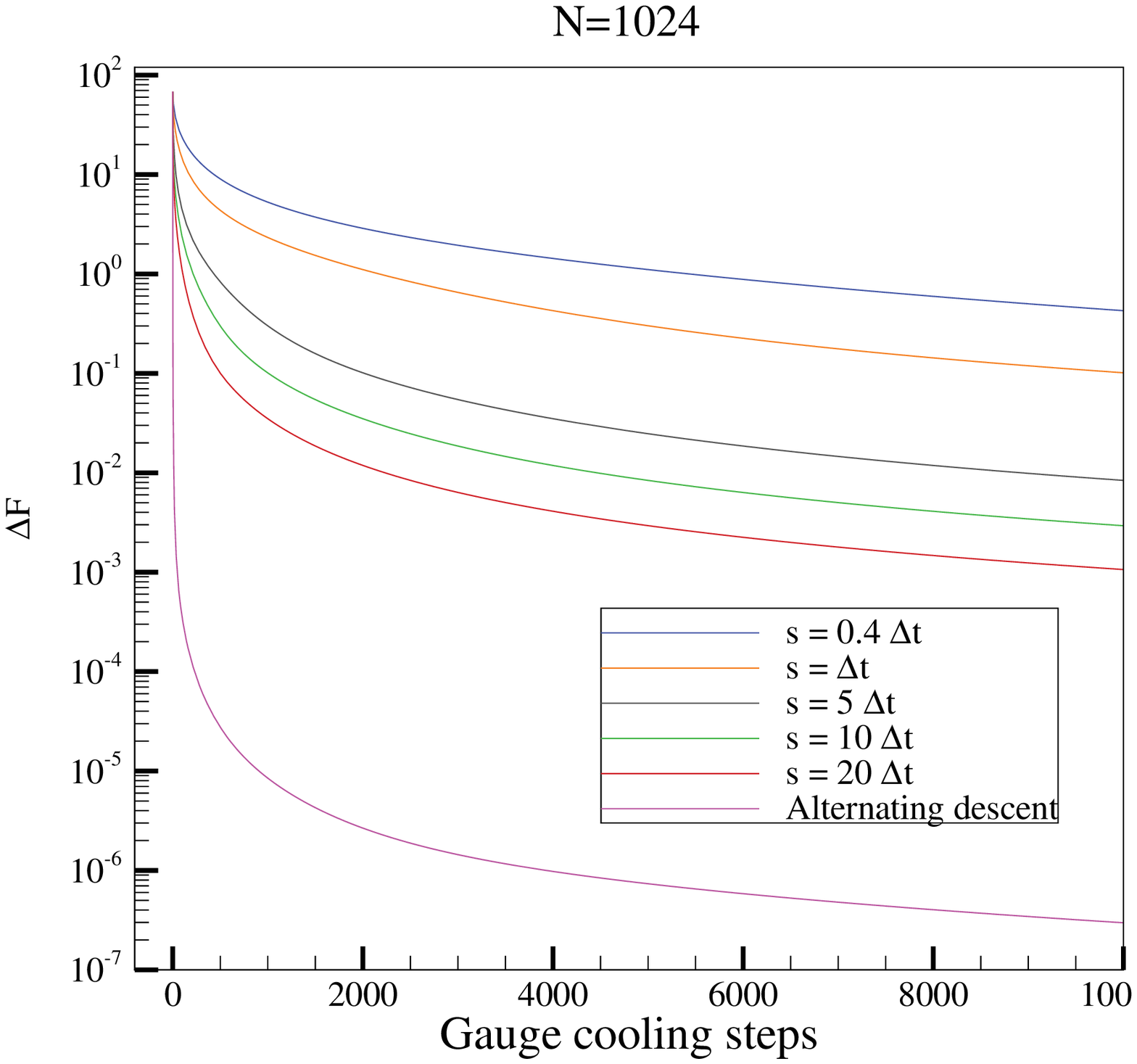}}
  \caption{The evolution of $\Delta F$ with gauge cooling steps increasing.}
  \label{fig:norm_s} 
\end{figure}

 To demonstrate the efficiency of the alternating descent method more clearly, we remove the complex Langevin dynamics and consider solving the optimization problem \eqref{eq:opt} directly. The link variables $\{U\}$ are set to be a random complexified gauge transformation applied to a random $SU(3)$ field, so that one of the solution to the problem \eqref{eq:opt} is the field before the gauge transformation, for which $\Delta F = 0$. Below we compare the performances of the alternating descent method and the gradient descent method with various step lengths $s$. The numbers of link variables are chosen to be $4, 32, 256$ and $1024$, and the results are plotted in Fig.\ref{fig:norm_s}. Note that the gradient descent method may get divergent when the time step is too large, and in our numerical test for every $N$, we have tried to maximize the step length $s$ which maintains the stability of iterative process. In all test cases, the alternating descent method shows its superiority in the cooling effect. In particular, it can efficiently bring down the value of $\Delta F$ by two to three orders of magnitude in the first few steps, and this effect is independent of the number of link variables. Such a property is highly desired in the simulation of lattice QCD, for it allows us to reduce the computational cost of gauge cooling.

\subsection{Heavy quark QCD}
In this example, we consider the QCD at finite chemical potential described in
\cite{Aarts2008}. Denoting any $x \in G$ by $x = (t, \mathbf{x})$, the fermion
determinant is given by
\begin{equation*}
\det M(\mu) = \prod_{\mathbf{x}} \det(I + C \mathcal{P}_\mathbf{x})^2 \det(I + C' \mathcal{P}_\mathbf{x}^{-1})^2
\end{equation*}
where $C = [2 \kappa \exp(\mu)^{N_0}]$, $C' = [2 \kappa \exp(-\mu)^{N_0}]$, and
\begin{equation*}
\mathcal{P}_\mathbf{x} = \prod_{t=1}^{N_0} U_{(t,\mathbf{x}),0}.
\end{equation*}
Define the plaquette
\begin{equation*}
U_{x,\mu \nu} = U_{x,\mu} U_{x+\hat{\mu},\nu} U_{x+\hat{\nu},\mu}^{-1} U_{x,\nu}^{-1}.
\end{equation*}
We can write the action as
\begin{equation*}
S(\{U\}) = - \ln (\det M) + S_B,
\end{equation*}
where
\begin{equation*}
S_B = -\beta \sum_{x \in G} \sum_{\mu<\nu} \left[ \frac{1}{6} (\tr U_{x,\mu \nu} + \tr U_{x,\mu \nu}^{-1}) - 1 \right].
\end{equation*}
We refer the readers to \cite{Aarts2008} for the Lie deriatives of this action. The same example has also been computed using the complex Langevin method in \cite{Pietri2007, Seiler2013, Fodor2015}.

In our simulation, we choose the chemical potential $\mu$ ranging from $0$ to $4.0$ with $\beta = 3.0$ and $\beta = 5.0$ to compute the expectation values of the following observables:
\begin{equation*}
\langle O \rangle = \frac{1}{3 N_1 N_2 N_3} \sum_{\mathbf{x}} \langle \tr \mathcal{P}_\mathbf{x} \rangle, \qquad \langle O^{-1} \rangle = \frac{1}{3 N_1 N_2 N_3} \sum_{\mathbf{x}} \langle \tr \mathcal{P}^{-1}_\mathbf{x} \rangle.
\end{equation*}
The value of $\kappa$ is set to be $0.12$. The lattice used in our numerical tests is $N_0 = N_1 = N_2 = N_3 = 4$, and the Langevin time step is $\Delta t = 2\times 10^{-5}$. Again, at every Langevin step, we perform three iterations for the gradient descent method and only one iteration for the alternating descent method. Since the lattice size in each direction is small, we expect that the gradient descent method and the alternating descent method have similar efficiency, while the alternating descent method still holds the advantage of its parameter-free property in each iteration. Moreover, averagely speaking, the gradient descent method spends $8.22$ms on gauge cooling, while the alternating descent method spends only $3.15$ms on gauge cooling.

For $\beta = 3.0$, we start to take samples every $50$ steps from $t = 2$ to $t = 10$ for all values of $\mu$. For $\beta = 5.0$, the Langevin dynamics becomes less stable for large $\mu$, and therefore we start $8$ parallel complex Langevin processes, and for each of them, we take one sample every $50$ steps from $t = 2$ to $t = 4$, so that in total, the same number of samples as the case $\beta = 3.0$ are collected to compute the obsersvables. 
\begin{figure}
  \centering
  \subfigure{
    \includegraphics[width=2.5in]{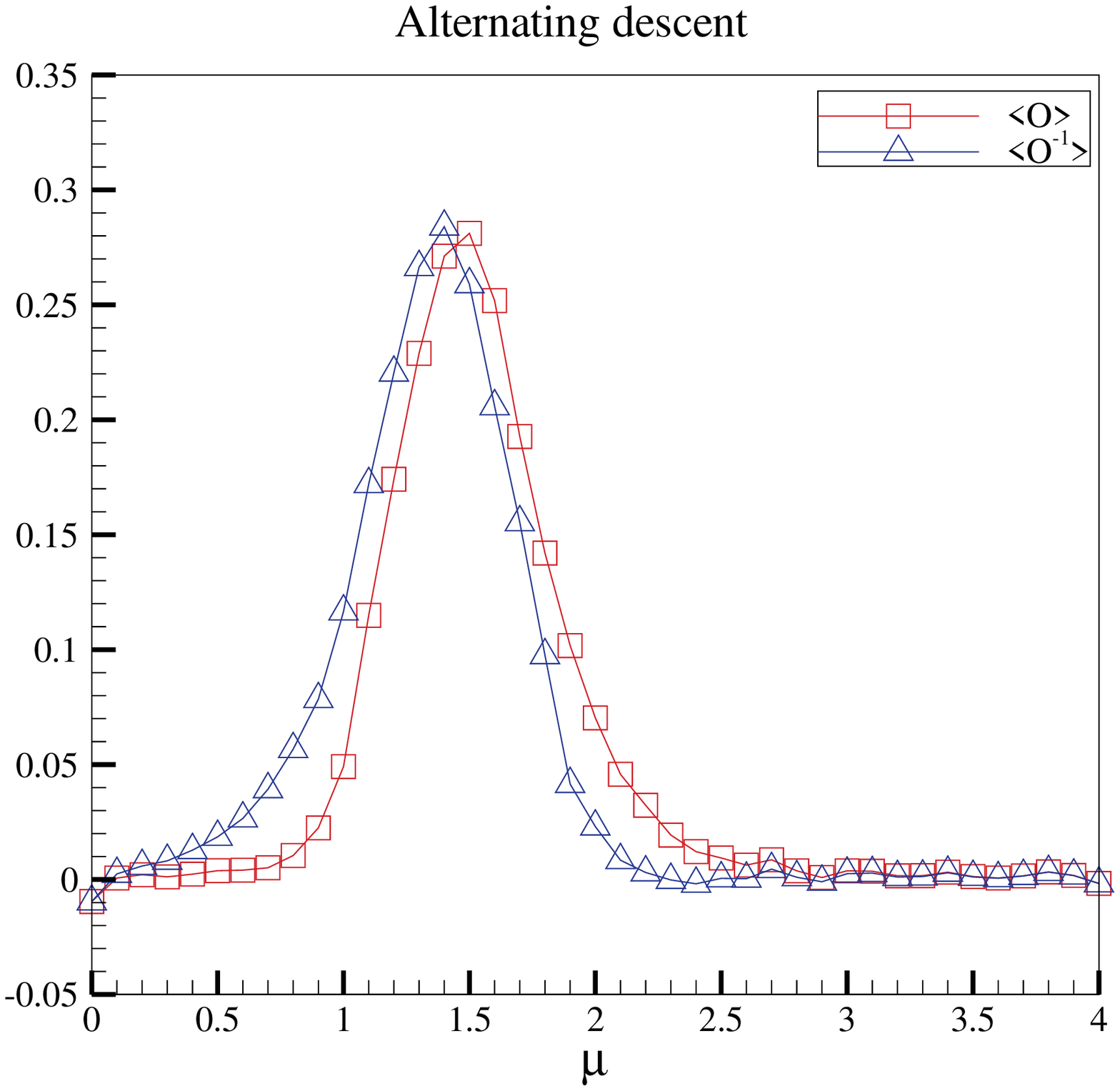}}
    \qquad
  \subfigure{
    \includegraphics[width=2.5in]{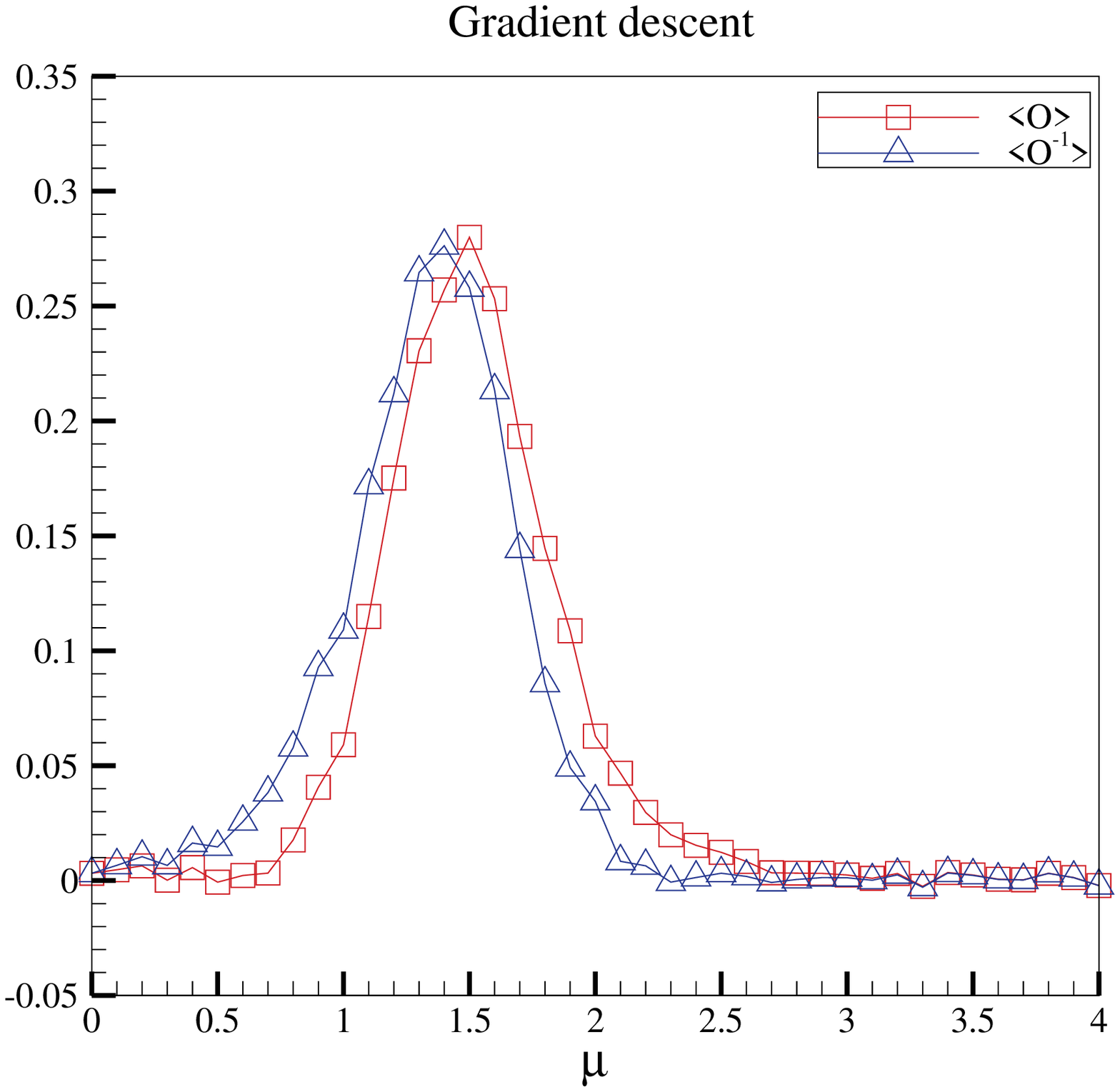}}
  \caption{The expectation values of observables with $\beta=3.0$.}
  \label{fig:obs_3} 
\end{figure} 
\begin{figure}
  \centering
  \subfigure{
    \includegraphics[width=2.5in]{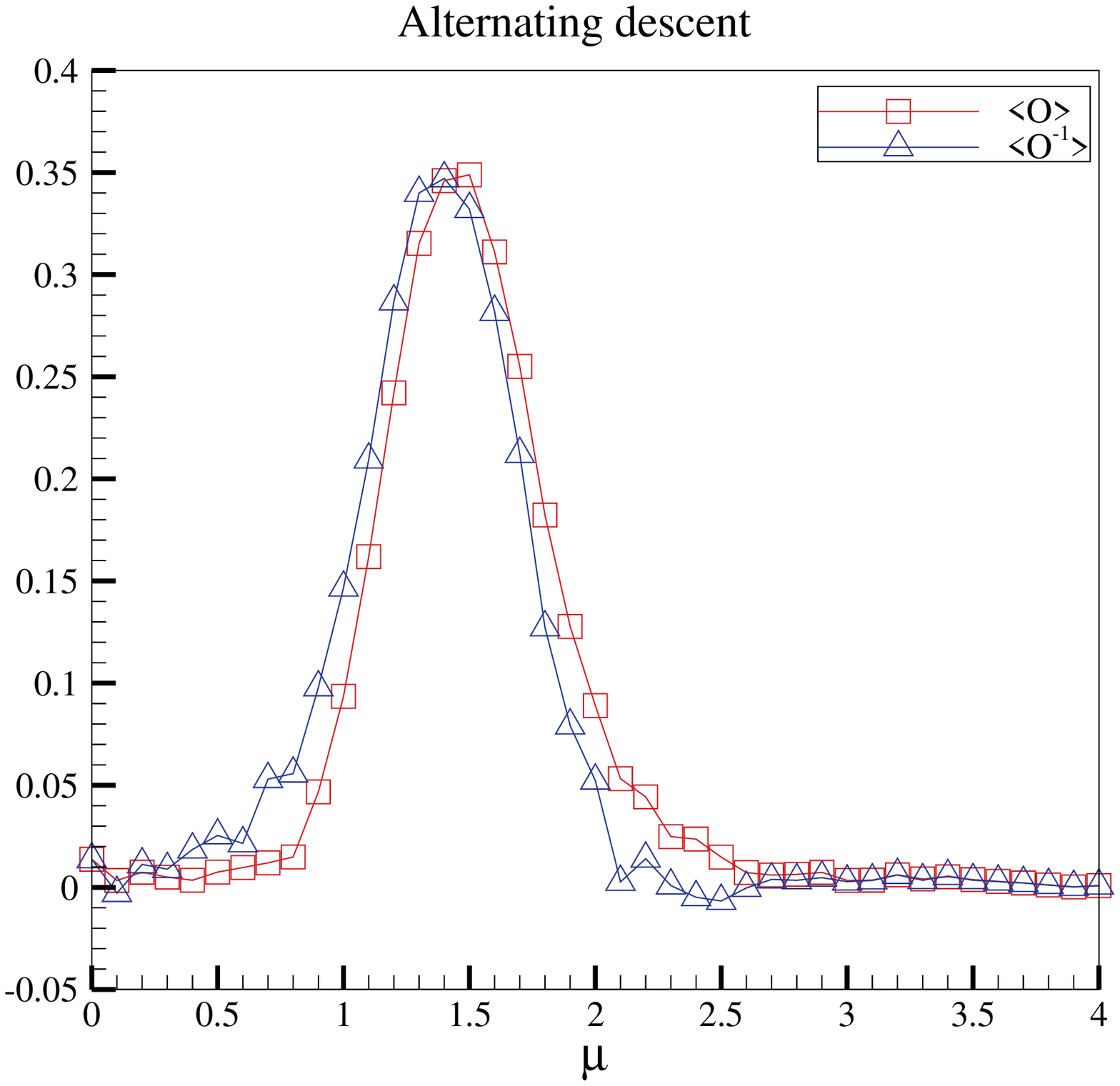}}
    \qquad
  \subfigure{
    \includegraphics[width=2.5in]{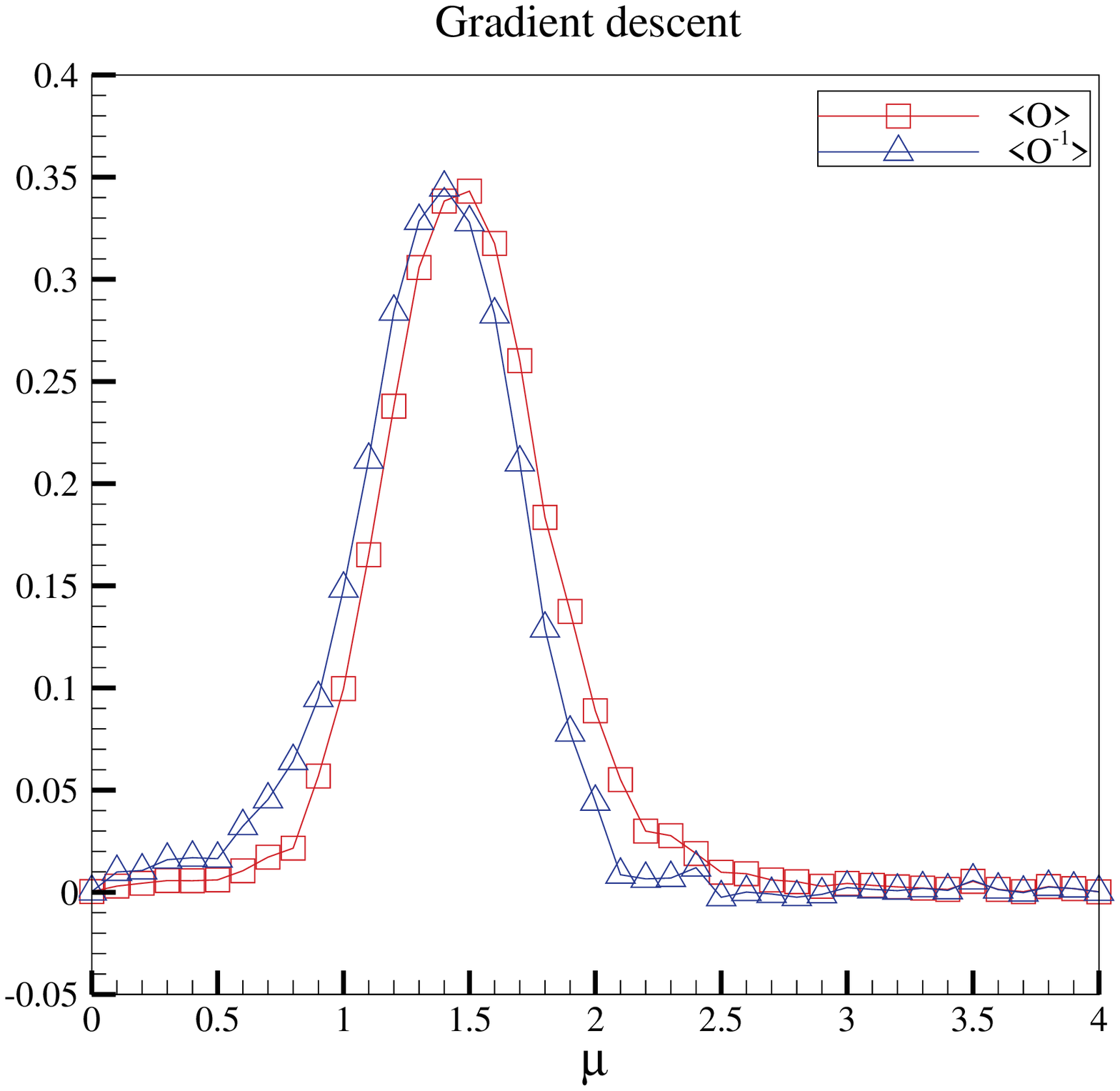}}
  \caption{The expectation values of observables with $\beta=5.0$.}
  \label{fig:obs_5} 
\end{figure} 

\begin{figure}
  \centering
  \subfigure{
    \includegraphics[width=2.5in]{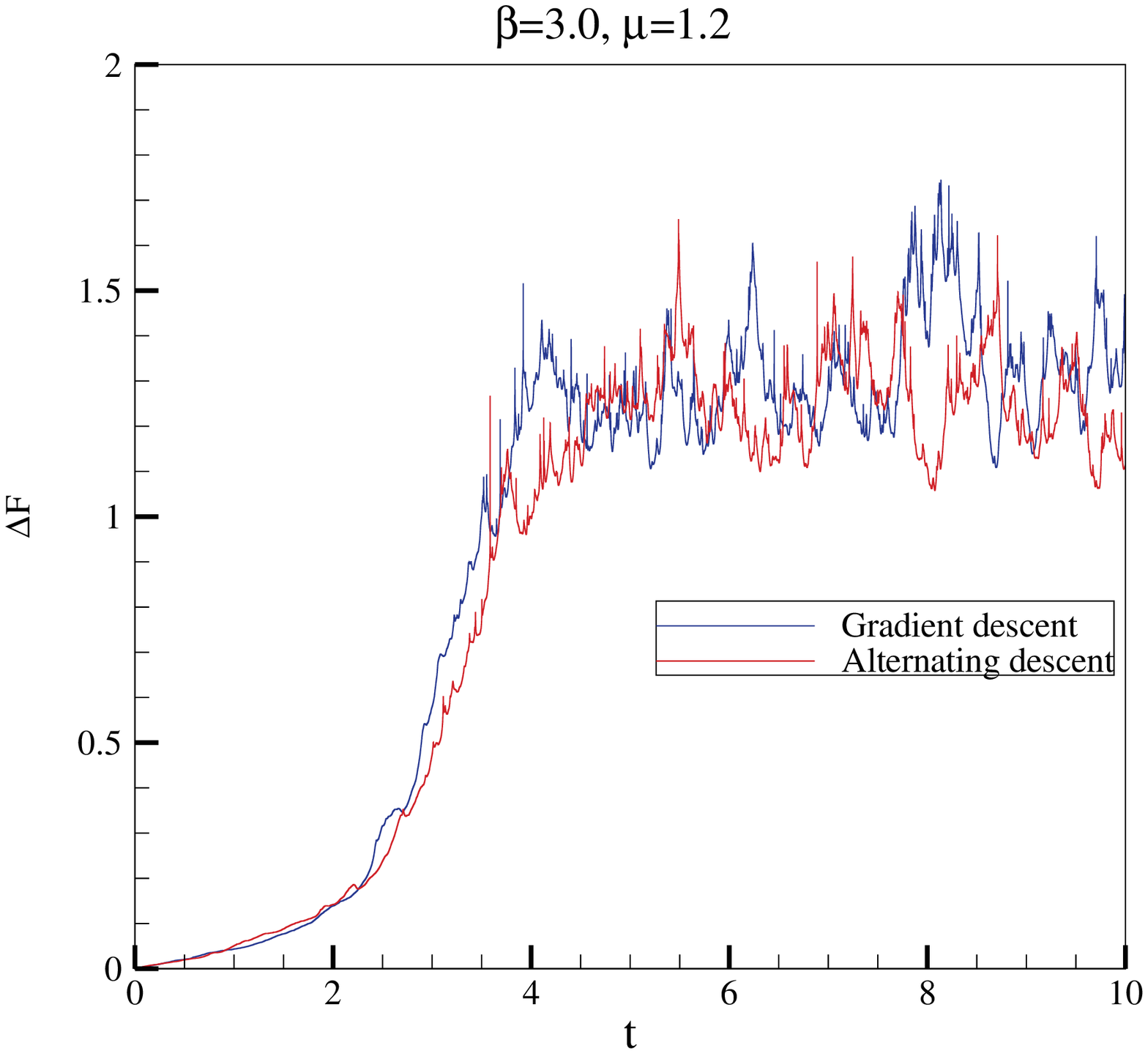}}
    \qquad
  \subfigure{
    \includegraphics[width=2.5in]{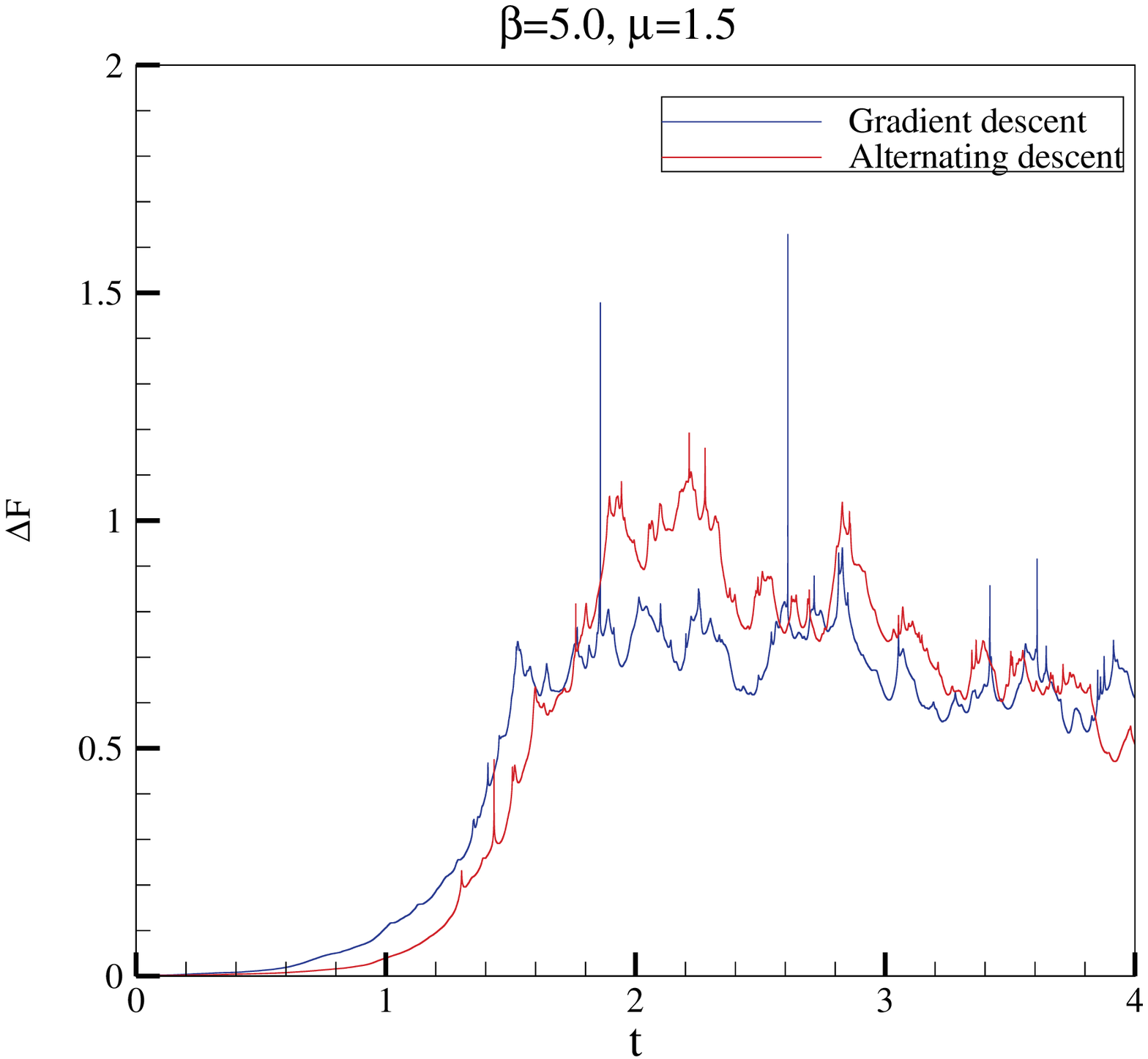}}
  \caption{The evolution of $\Delta F$ with different gauge cooling techniques.}
  \label{fig:norm_2} 
\end{figure}

The observables obtained in both methods are plotted in Fig.\ref{fig:obs_3} and Fig.\ref{fig:obs_5}. For small $\mu$, these results can be cross-checked with the analytical results in \cite{Pietri2007}, and the bell-shaped curves agree with the results computed in \cite{Seiler2013}. As expected, the results of both methods well agree with each other, while it is worth mentioning that for the gradient descent method, three iterations are applied to solve the optimization problem after every time step, while for the alternating descent method, only one iteration is applied. Fig.\ref{fig:norm_2} shows the evolution of $\Delta F$. Our one-dimensional examples in Section \ref{sec:1D} suggest similar behavior of two methods, as is generally true in our numerical tests. However, it can be observed in the case $\beta = 5.0$ and $\mu = 1.5$ that the gradient descent method shows several sharp peaks of the norm, indicating the failure of gauge cooling in certain particular cases. While the details are still to be further studied, it shows better stability of the alternating descent method. We also expect better performance of the alternating descent method when the lattice size gets larger.

%% file: Conclusion.tex
\section{Conclusion} \label{sec:conclusion}
We have introduced a new gauge cooling technique called alternating descent method, which helps effectively find the gauge transformation that pulls the complexified gauge field closer to the unitary field. The key properties of this method include:
\begin{itemize}
\item No parameter needs to be chosen in each iteration.
\item It is easy to implement since the cooling step is implemented site by site.
\item The objective function declines fast especially in the first few steps.
\item The performance of the method is stable for various numbers of link variables.
\end{itemize}
Our numerical tests show that when the number of link variables is small, its performance is comparable to the gradient descent method with carefully chosen step length; and when the lattice size gets larger, the alternating descent method gets more superior. The underlying mechanism of the alternating descent method needs to be further studied, and we expect more applications to be carried out in our future work.